\documentclass[twocolumn,showpacs]{revtex4}

\usepackage{graphicx}

\newcommand{\SK}[0]{Super-Kamiokande}

\begin{document}

\title{Search for Supernova Relic Neutrinos at Super-Kamiokande}

\newcounter{foots}

\newcommand{\authoraticrr}{$^{1}$}
\newcommand{\authoratbu}{$^{2}$}
\newcommand{\authoratbnl}{$^{3}$}
\newcommand{\authoratuci}{$^{4}$}
\newcommand{\authoratcsu}{$^{5}$}
\newcommand{\authoratgmu}{$^{6}$}
\newcommand{\authoratgifu}{$^{7}$}
\newcommand{\authoratuh}{$^{8}$}
\newcommand{\authoratkek}{$^{9}$}
\newcommand{\authoratkobe}{$^{10}$}
\newcommand{\authoratkyoto}{$^{11}$}
\newcommand{\authoratlanl}{$^{12}$}
\newcommand{\authoratlsu}{$^{13}$}
\newcommand{\authoratumd}{$^{14}$}
\newcommand{\authoratmit}{$^{15}$}
\newcommand{\authoratduluth}{$^{16}$}
\newcommand{\authoratsuny}{$^{17}$}
\newcommand{\authoratnagoya}{$^{18}$}
\newcommand{\authoratniigata}{$^{19}$}
\newcommand{\authoratosaka}{$^{20}$}
\newcommand{\authoratseoul}{$^{21}$}
\newcommand{\authoratseika}{$^{22}$}
\newcommand{\authoratshizuoka}{$^{23}$}
\newcommand{\authorattohoku}{$^{24}$}
\newcommand{\authorattokyo}{$^{25}$}
\newcommand{\authorattokai}{$^{26}$}
\newcommand{\authorattit}{$^{27}$}
\newcommand{\authoratwarsaw}{$^{28}$}
\newcommand{\authoratuw}{$^{29}$}

\newcommand{\addressoficrr}[1]{$^{1}$ #1 }
\newcommand{\addressofbu}[1]{$^{2}$ #1 }
\newcommand{\addressofbnl}[1]{$^{3}$ #1 }
\newcommand{\addressofuci}[1]{$^{4}$ #1 }
\newcommand{\addressofcsu}[1]{$^{5}$ #1 }
\newcommand{\addressofgmu}[1]{$^{6}$ #1 }
\newcommand{\addressofgifu}[1]{$^{7}$ #1 }
\newcommand{\addressofuh}[1]{$^{8}$ #1 }
\newcommand{\addressofkek}[1]{$^{9}$ #1 }
\newcommand{\addressofkobe}[1]{$^{10}$ #1 }
\newcommand{\addressofkyoto}[1]{$^{11}$ #1 }
\newcommand{\addressoflanl}[1]{$^{12}$ #1 }
\newcommand{\addressoflsu}[1]{$^{13}$ #1 }
\newcommand{\addressofumd}[1]{$^{14}$ #1 }
\newcommand{\addressofmit}[1]{$^{15}$ #1 }
\newcommand{\addressofduluth}[1]{$^{16}$ #1 }
\newcommand{\addressofsuny}[1]{$^{17}$ #1 }
\newcommand{\addressofnagoya}[1]{$^{18}$ #1 }
\newcommand{\addressofniigata}[1]{$^{19}$ #1 }
\newcommand{\addressofosaka}[1]{$^{20}$ #1 }
\newcommand{\addressofseoul}[1]{$^{21}$ #1 }
\newcommand{\addressofseika}[1]{$^{22}$ #1 }
\newcommand{\addressofshizuoka}[1]{$^{23}$ #1 }
\newcommand{\addressoftohoku}[1]{$^{24}$ #1 }
\newcommand{\addressoftokyo}[1]{$^{25}$ #1 }
\newcommand{\addressoftokai}[1]{$^{26}$ #1 }
\newcommand{\addressoftit}[1]{$^{27}$ #1 }
\newcommand{\addressofwarsaw}[1]{$^{28}$ #1 }
\newcommand{\addressofuw}[1]{$^{29}$ #1 }

\author{
{\large The Super-Kamiokande Collaboration} \\
\bigskip
%
M.~Malek\authoratsuny,
M.~Morii\authorattit,
%
S.~Fukuda\authoraticrr,
Y.~Fukuda\authoraticrr,
M.~Ishitsuka\authoraticrr,
Y.~Itow\authoraticrr,
T.~Kajita\authoraticrr,
J.~Kameda\authoraticrr,
K.~Kaneyuki\authoraticrr,
K.~Kobayashi$^{1,17}$
% ...
\\
\smallskip
\smallskip
\footnotesize
\it
\addressoficrr{Institute for Cosmic Ray Research, University of Tokyo, Kashiwa, Chiba 277-8582, Japan}\\
\addressofbu{Department of Physics, Boston University, Boston, MA 02215, USA}\\
\addressofbnl{Physics Department, Brookhaven National Laboratory, Upton, NY 11973, USA}\\
\addressofuci{Department of Physics and Astronomy, University of California, Irvine, Irvine, CA 92697-4575, USA }\\
\addressofcsu{Department of Physics, California State University, Dominguez Hills, Carson, CA 90747, USA}\\
\addressofgmu{Department of Physics, George Mason University, Fairfax, VA 22030, USA }\\
\addressofgifu{Department of Physics, Gifu University, Gifu, Gifu 501-1193, Japan}\\
\addressofuh{Department of Physics and Astronomy, University of Hawaii, Honolulu, HI 96822, USA}\\
\addressofkek{Institute of Particle and Nuclear Studies, High Energy Accelerator Research Organization (KEK), Tsukuba, Ibaraki 305-0801, Japan }\\
\addressofkobe{Department of Physics, Kobe University, Kobe, Hyogo 657-8501, Japan}\\
\addressofkyoto{Department of Physics, Kyoto University, Kyoto 606-8502, Japan}\\
\addressoflanl{Physics Division, P-23, Los Alamos National Laboratory, Los Alamos, NM 87544, USA }\\
\addressoflsu{Department of Physics and Astronomy, Louisiana State University, Baton Rouge, LA 70803, USA }\\
\addressofumd{Department of Physics, University of Maryland, College Park, MD 20742, USA }\\
\addressofmit{Department of Physics, Massachusetts Institute of Technology, Cambridge, MA 02139, USA}\\
\addressofduluth{Department of Physics, University of Minnesota, Duluth, MN 55812-2496, USA}\\
\addressofsuny{Department of Physics and Astronomy, State University of New York, Stony Brook, NY 11794-3800, USA}\\
\addressofnagoya{Department of Physics, Nagoya University, Nagoya, Aichi 464-8602, Japan}\\
\addressofniigata{Department of Physics, Niigata University, Niigata, Niigata 950-2181, Japan }\\
\addressofosaka{Department of Physics, Osaka University, Toyonaka, Osaka 560-0043, Japan}\\
\addressofseoul{Department of Physics, Seoul National University, Seoul 151-742, Korea}\\
\addressofseika{International and Cultural Studies, Shizuoka Seika College, Yaizu, Shizuoka, 425-8611, Japan}\\
\addressofshizuoka{Department of Systems Engineering, Shizuoka University, Hamamatsu, Shizuoka 432-8561, Japan}\\
\addressoftohoku{Research Center for Neutrino Science, Tohoku University, Sendai, Miyagi 980-8578, Japan}\\
\addressoftokyo{The University of Tokyo, Tokyo 113-0033, Japan }\\
\addressoftokai{Department of Physics, Tokai University, Hiratsuka, Kanagawa 259-1292, Japan}\\
\addressoftit{Department of Physics, Tokyo Institute for Technology, Meguro, Tokyo 152-8551, Japan }\\
\addressofwarsaw{Institute of Experimental Physics, Warsaw University, 00-681 Warsaw, Poland }\\
\addressofuw{Department of Physics, University of Washington, Seattle, WA 98195-1560, USA}\\
}

\affiliation{ } 



\begin{abstract}

A search for the relic neutrinos from all past core-collapse supernovae was conducted using 1496 days of data from the \SK\, detector.  This analysis looked for electron-type anti-neutrinos that had produced a positron with an energy greater than 18~MeV.  In the absence of a signal, 90\% C.L. upper limits on the total flux were set for several theoretical models; these limits ranged from 20 to 130~$\bar{\nu}_e$~cm$^{-2}$~s$^{-1}$.  Additionally, an upper bound of $1.2\; \bar{\nu}_e$~cm$^{-2}$~s$^{-1}$ was set for the supernova relic neutrino flux in the energy region $E_\nu > 19.3$~MeV.

\end{abstract}

\pacs{95.85.Ry,97.60.Bw,14.60.Lm,96.40.Tv}
%

\maketitle

\renewcommand{\baselinestretch}{1.1}


During a core-collapse supernova, approximately $10^{53}$ ergs of energy are released, about 99\% of which are in the form of neutrinos.  To date, the only time that a burst of such neutrinos has been detected was in the case of SN1987A~\cite{1987-KAM,1987-IMB}.  However, it is generally believed that core-collapse supernovae have occurred throughout the universe since the formation of stars.  Thus, there should exist a diffuse background of neutrinos originating from all the supernovae that have ever occurred.  Detection of these supernova relic neutrinos (SRN) would offer insight about the history of star formation and supernovae explosions in the universe.  


All types of neutrinos and anti-neutrinos are emitted from a core-collapse supernova, but not all are equally detectable.  The $\bar{\nu}_e$ is most likely to be detected by Super-Kamiokande (SK).  It interacts primarily through inverse $\beta$ decay ($\bar{\nu}_e\;p \rightarrow n\;e^+$, $E_e$ = $E_\nu$ - 1.3~MeV) with a cross section that is two orders of magnitude greater than that of neutrino-electron elastic scattering.  All further discussion herein of the SRN refers only to the $\bar{\nu}_e$. 


Several methods have been used to model the SRN flux and spectrum~\cite{TS-1995,TSY-1996,malaney,hartmann,KSW,LMA}, with flux predictions ranging from 2 -- 54~cm$^{-2}$~s$^{-1}$.  In this paper, SK search results are compared to SRN predictions based on a galaxy evolution model~\cite{TSY-1996}, a cosmic gas infall model~\cite{malaney}, cosmic chemical evolution studies~\cite{hartmann}, and observations of heavy metal abundances~\cite{KSW}.  A model that assumes a constant supernova rate~\cite{TSY-1996} was also considered; this model was used to set the previous SRN flux limit at Kamiokande-II~\cite{hirata}.

It has been shown that neutrinos undergo flavor oscillation~\cite{sk-atm-osc,sk-hep-1260,sno-2001}.  Therefore, an SRN spectrum that includes the effects of neutrino mixing was also considered~\cite{LMA}.  All six species of neutrinos are emitted during a core-collapse supernova.  However, the $\bar{\nu}_\mu$ and the $\bar{\nu}_\tau$ decouple from the neutrinosphere earlier than the $\bar{\nu}_e$, resulting in a higher temperature for these flavors; thus, neutrino mixing would harden the $\bar{\nu}_e$ energy spectrum.  The cross section for inverse $\beta$ decay increases as the square of the $\bar{\nu}_e$ energy, so oscillation would enhance the SRN signal.  In this paper, only a large mixing angle MSW solution (LMA) was used to distort the SRN spectrum -- the LMA solution is also favored by the available solar neutrino data~\cite{skglob,snoglob}.




This paper presents the results of a search for SRNs in the Super-Kamiokande detector.  SK is a water Cherenkov detector, with a fiducial mass of 22.5~kton, located in the Kamioka Mine in Gifu, Japan.  Descriptions of the detector can be found elsewhere~\cite{linac}.  The data reported here were collected between May 31, 1996, and July 15, 2001, yielding a total SRN search livetime of 1496 days.  Backgrounds to the SRN signal are solar neutrinos, atmospheric neutrinos, and muon-induced spallation products.  Background reduction takes place in the following steps:  spallation cut, sub-event cut, Cherenkov angle cut, and solar direction cut.  

Spallation is the most serious background, and the ability to remove it determines the lower threshold of the SRN search.  Spallation products are also relevant to solar neutrino studies, and so a likelihood function had been developed that uses information about the muons preceding the possible spallation event~\cite{sk-hep-1260}.  To permit a low analysis threshold for the SRN search, a tighter spallation cut was implemented:  in addition to the likelihood function cut, all events that occur less than 0.15~seconds after a cosmic ray muon are rejected.  The spallation cut is applied to all events reconstructed with $\rm E < 34$~MeV, and introduces a deadtime of 36\%.  No discernible spallation events with energies above 18~MeV remain in the data after this cut is applied and so 18~MeV was set as the lower analysis threshold.





The sub-event cut removes muons with kinetic energy (T) in the range of 50 -- 140~MeV.  Cosmic ray muons, which have much higher energies, originate outside of SK and are removed because they produce a veto signal in the SK outer detector (OD).  However, atmospheric $\nu_\mu$ can produce muons within the inner detector (ID) via charged current quasi-elastic scattering ($\nu_\mu\;N \rightarrow \mu\;N'$); such muons will not have been tagged by the OD, but will be visible in the ID.  Muons with low energies will stop in SK and produce a decay electron; often the muon and decay electron are found in the same event.  When this happens, the decay electron is referred to as a ``sub-event."  After the vertex of each event was found and the flight time of the Cherenkov photons was subtracted, the event's $1.3\, \mu$s time window was searched; if more than one timing peak was present, then the event was removed.  The sub-event cut was tested on simulated muons in the relevant energy range, and it was shown to remove 34\% of the muon background.

The remaining muons are removed by the Cherenkov angle cut.  This cut exploits the mass difference between the muon and the positron, which results in a difference in their Cherenkov angles $\theta_C$.  Positrons with $\rm E > 18~MeV$ have $\theta_C$ $\approx 42$~degrees.  Muons with $\rm T < 140~MeV$ have $\rm \theta_C < 34$~degrees; thus, all particles with $\theta_C$ $< 37$~degrees were removed from the data.  The efficiency of this selection criterion for retaining signal is 98\%, as determined by applying it to simulated positron events.  Using simulated muon events, it was shown that applying the Cherenkov angle cut and the sub-event cut together results in the rejection of 96\% of the muon background.  Furthermore, it was found that the full reduction removes $> 99\%$ of the muons.  The Cherenkov angle cut was also used to remove events with $\theta_C > 50$~degrees; this eliminated events without clear Cherenkov rings, such as multiple $\gamma$~rays emitted during a nuclear de-excitation, from the data sample.

Finally, a cut on the direction of the event is made to remove contamination from solar neutrinos.  Events with $\rm E < 34$~MeV were removed if the reconstructed event direction was within thirty degrees of the vector from the Sun to the Earth at the time of the event.

\begin{figure}[bt]
\center{\includegraphics[width=8cm,clip]{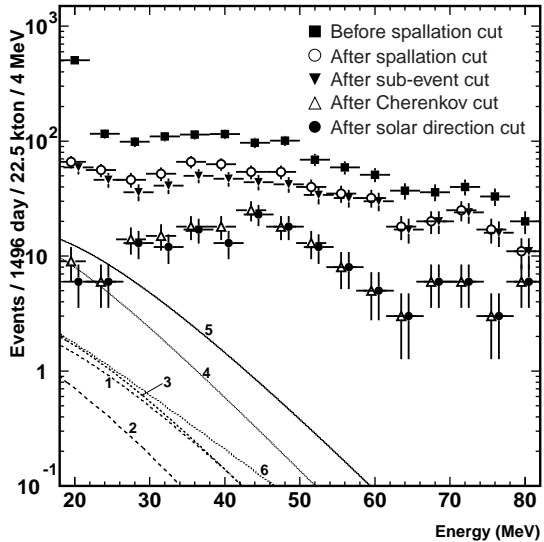}}
\caption{Energy spectrum at each reduction step.  In the final data set, the spallation cut and solar direction cut are only applied in the first four bins.  The numbered lines represent the corresponding theoretical predictions from Table~\ref{tab:results}.}
\label{reduc}
\end{figure}

By simulating positrons created from SRN, it was found that the efficiency of the full data reduction is $47 \pm 0.4\%$ for $E \le 34$~MeV, and $79 \pm 0.5\%$ for $E > 34$~MeV.  Figure~\ref{reduc} plots the energy spectrum after each cut.



After applying the selection criteria, two irreducible backgrounds remain.  The first is atmospheric $\nu_e$ and $\bar{\nu}_e$ events.  The second comes from atmospheric $\nu_\mu$ that interact to form a muon with $\rm T < 50~MeV$.  The energy of these muons is below the threshold for emitting Cherenkov photons, so they are said to be invisible.  Decay electrons from visible muons can be eliminated; however, when an invisible muon decays there is no way to tag the resulting electron as a background event.


The energy spectra of these backgrounds have shapes that are very different from each other and from the SRN signal shape.  In the region where SRN events are expected (18 -- 34~MeV), the dominant background is decay electrons from invisible muons, which have energies that are distributed according to the Michel spectrum.  However, at higher energies, atmospheric $\nu_e$ events distort the Michel spectrum.  To evaluate the distortion, it is necessary to extend the upper analysis threshold to energies where only atmospheric $\nu_e$ events are present.  Decay electrons have a maximum energy of 53~MeV, but may be detected up to $\sim$ 65~MeV due to the energy resolution of SK.  Beyond 65~MeV, only atmospheric $\nu_e$ are found, so the upper analysis threshold was set above 65~MeV and the data were analyzed with a three parameter fit.  

\begin{figure}
\center{\includegraphics[width=8cm,clip]{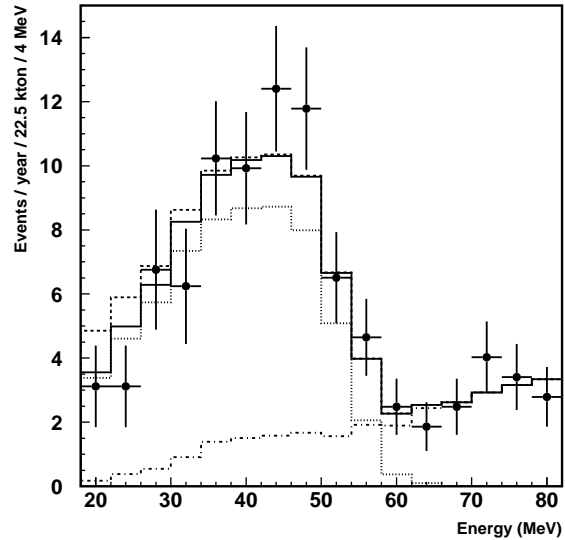}}
\caption{Energy spectrum of SRN candidates.  The dotted and dash-dot histograms are the fitted backgrounds from invisible muons and atmospheric $\nu_e$.  The solid histogram is the sum of these two backgrounds.  The dashed line shows the sum of the total background and the 90\% upper limit of the SRN signal.}
\label{fit}
\end{figure}

To determine the final shape of the backgrounds, 100 years of simulated events were generated per background.  The initial shape of the decay electrons was determined by the Michel spectrum; the initial shape of the atmospheric $\nu_e$ events was obtained from previous works~\cite{sub-gev,barr}.  The background simulations were subjected to the full reduction, and the shape of the resulting spectra were used to fit the data; each of the SRN models was treated similarly.  For the fitting, the data were divided into sixteen energy bins, each 4~MeV wide (see Figure~\ref{fit}), and the following ${\chi}^2$ function was minimized with respect to $\alpha$, $\beta$, and $\gamma$.

{\small
\begin{equation}\label{eqn:csq}
{\chi^2} =  \sum_{l = 1}^{16}
  \frac{[(\alpha \cdot A_l) + (\beta \cdot B_l) + (\gamma \cdot C_l) - N_l]^2}{\sigma_{stat}^2 + \sigma_{sys}^2}
\end{equation}
}

\begin{table*}
\caption{\rm The SRN search results are presented for six theoretical models.  The first column describes the method used to calculate the SRN flux.  The second column shows the efficiency-corrected limit on the SRN event rate at SK.  The third column is the flux limit set by SK, which can be compared with the theoretical predictions that are shown in the fourth column.  The fifth column shows the flux predictions above a threshold of $\rm E_\nu > 19.3$~MeV .  Note that the heavy metal abundance calculation only sets a theoretical upper bound on the SRN flux~\cite{KSW}.}  
\label{tab:results}
\begin{center}
\begin{tabular}{ccccc}
\hline
\hline
  Theoretical model          & Event rate limit & SRN flux limit   & Predicted flux & Predicted flux \\
                             &   (90\% C.L.)    &  (90\% C.L.)     &                & ($\rm E_\nu > 19.3$~MeV) \\
\hline
Galaxy evolution~\cite{TSY-1996}        &   $<$ 3.2 events/year        & $<$ 130 $\bar{\nu}_e$~cm$^{-2}$~s$^{-1}$ &   44 $\bar{\nu}_e$~cm$^{-2}$~s$^{-1}$ & 0.41 $\bar{\nu}_e$~cm$^{-2}$~s$^{-1}$ \\

Cosmic gas infall~\cite{malaney}  &   $<$ 2.8 events/year        &  $<$ 32 $\bar{\nu}_e$~cm$^{-2}$~s$^{-1}$ & $ 5.4\; \bar{\nu}_e$~cm$^{-2}$~s$^{-1}$ &  0.20 $\bar{\nu}_e$~cm$^{-2}$~s$^{-1}$  \\

Cosmic chemical evolution~\cite{hartmann}  &   $<$ 3.3 events/year        &  $<$ 25 $\bar{\nu}_e$~cm$^{-2}$~s$^{-1}$ & $ 8.3\; \bar{\nu}_e$~cm$^{-2}$~s$^{-1}$ &  0.39 $\bar{\nu}_e$~cm$^{-2}$~s$^{-1}$  \\

Heavy metal abundance~\cite{KSW}   &   $<$ 3.0 events/year        &  $<$ 29 $\bar{\nu}_e$~cm$^{-2}$~s$^{-1}$ & $< 54\; \bar{\nu}_e$~cm$^{-2}$~s$^{-1}$ & $<$ 2.2 $\bar{\nu}_e$~cm$^{-2}$~s$^{-1}$  \\

Constant supernova rate~\cite{TSY-1996} &   $<$ 3.4 events/year        & $<$ 20 $\bar{\nu}_e$~cm$^{-2}$~s$^{-1}$ &   52 $\bar{\nu}_e$~cm$^{-2}$~s$^{-1}$ & 3.1 $\bar{\nu}_e$~cm$^{-2}$~s$^{-1}$ \\

Large mixing angle osc.~\cite{LMA} &   $<$ 3.5 events/year        &  $<$ 31 $\bar{\nu}_e$~cm$^{-2}$~s$^{-1}$ &   11 $\bar{\nu}_e$~cm$^{-2}$~s$^{-1}$ & 0.43 $\bar{\nu}_e$~cm$^{-2}$~s$^{-1}$ \\
\hline
\hline
\end{tabular}
\end{center}
\end{table*}

In this equation, the sum $l$ is over all energy bins and $N_l$ is the number of events in the $l^{th}$ bin.  $A_l$, $B_l$, and $C_l$ represent, respectively, the fractions of the SRN, Michel, and atmospheric $\nu_e$ spectra that are in the $l^{th}$ bin.  $\alpha$, $\beta$, and $\gamma$ are the fitting parameters for the number of SRN events, decay electrons, and atmospheric $\nu_e$ events.

The total number of events in the data sample is small, so the statistical error $\sigma_{stat}$ is the dominant term in the denominator.  The systematic error $\sigma_{sys}$ considers the effects that uncertainties in the spectrum shapes have on the SRN result.  Such uncertainties originate from the reduction, the SK energy resolution, the theoretical atmospheric $\nu_e$ spectrum, and other sources.  For all bins, $\sigma_{sys} \approx$ 6\%, which is always much smaller than $\sigma_{stat}$.

The efficiency-corrected event rate spectrum of SRN candidates and the results of the fit are shown in Figure~\ref{fit}.  The best fits to $\gamma$ and $\beta$ are indicated, respectively, by the dot-dashed and dotted lines.  The solid line is the sum of these lines and represents the total background.  For all six models, the best fit to $\alpha$ was zero and the minimum ${\chi}^2$ value was 8.1 for 13 degrees of freedom.  A 90\% C.L. limit on $\alpha$ was set for each model; the dashed line represents the sum of the background and the upper bound on $\alpha$ for the galaxy evolution model.  This line shows the type of distortion in the Michel spectrum that would be indicative of an SRN signal.  

Figure~\ref{fit} shows that the expected backgrounds fit the data well.  In this analysis, no flux normalization was chosen for the background rates; only the shapes were used.  This is because there are large uncertainties ($\sim$~30\%) in the atmospheric neutrino fluxes at these very low energies.  As a consistency check, the fit results for the number of background events were compared to the predictions, which were determined by applying the reduction cuts to 100 years of simulated background events and normalizing for livetime.  For 1496 days of data, the expected number of atmospheric $\nu_e$ events is $75 \pm 23$, which is consistent with the best fit result of $88 \pm 12$ events.  To determine the expected number of decay electrons from invisible muons, neutrino oscillation must be considered~\cite{sk-atm-osc}; given the low energy of the atmospheric $\nu_\mu$ that produce invisible muons, it is assumed that half of the $\nu_\mu$ have oscillated into $\nu_\tau$.  With this assumption, the predicted number of decay electron events is $145 \pm 43$, which is consistent with the best fit result of $174 \pm 16$ events.
	
The limit on $\alpha$ can be used to derive a 90\% C.L. limit on the SRN flux from each model.  The number of SRN events is related to the total flux $F$ by the following equation:

\begin{equation}
\label{e:f}
F = \frac{\alpha}{N_p \times \tau \int_{\rm 19.3\, MeV}^{\infty}{f(E_\nu)\sigma(E_\nu)\epsilon(E_\nu)dE_\nu}}
\end{equation}

In this equation, $N_p$ is the number of free protons in SK ($\rm 1.5 \times 10^{33}$), $\tau$ is the detector livetime (1496~days), $\epsilon(E)$ is the signal detection efficiency, $\sigma(E)$ is the cross section for the inverse $\beta$ decay ($\rm 9.52 \times 10^{-44}~E_e~p_e$), and $f(E)$ is the normalized SRN spectrum shape.  The integral spans the energy range of the neutrinos that produce positrons in the observed region.




Using the above values, the 90\% C.L. SRN flux limit was calculated for each model.  The results are in the third column of Table~\ref{tab:results}, and can be compared with the predictions, which are in the fourth column.  For the galaxy evolution model~\cite{TSY-1996}, the cosmic gas infall model~\cite{malaney}, and the cosmic chemical evolution model~\cite{hartmann}, the SK limits are larger than the predictions by a factor of three to six.  In these models, the dominant contribution to the SRN flux comes from supernovae in the early universe, so the neutrino energy is red-shifted below the 18~MeV threshold.  The heavy metal abundance model primarily considers supernovae at red-shifts $z < 1$, so SK is sensitive to more of the SRN flux.  For this model, the flux limit is smaller than the calculated total flux~\cite{KSW}.  However, this prediction is only a theoretical upper limit, so these results can constrain this model but they cannot eliminate it.  The LMA model~\cite{LMA} has a harder energy spectrum, and so SK is sensitive to a larger fraction of the SRN flux.  The increased sensitivity is offset by the fact that this hardened spectrum also results in a larger limit for $\alpha$; thus, the SK flux limit is still nearly a factor of three larger than the prediction.

The total SRN flux predicted by the constant model scales with the rate of core-collapse supernovae, and so the SRN flux limit can be used to set a 90\% C.L. upper limit on the constant supernova rate.  The SRN flux prediction quoted in this paper is based on a reasonable supernova rate of $\rm 1.6~\times 10^3$~SN~year$^{-1}$~Mpc$^{-3}$.  The observed SRN flux limit ($20\; \bar{\nu}_e$~cm$^{-2}$~s$^{-1}$) corresponds to a supernova rate limit of $6.2 \times 10^2$~SN~year$^{-1}$~Mpc$^{-3}$.  Thus, the constant model can be ruled out, as the limit on the supernova rate is too low to be consistent with the observed abundance of oxygen~\cite{oxygen,TSY-1996}, which is synthesized within the massive stars that become supernovae.  At Kamiokande-II, a flux limit of $780\; \bar{\nu}_e$~cm$^{-2}$~s$^{-1}$ was set with the assumption of a constant supernova model~\cite{hirata}; the SK limit is 39 times more stringent.

The SRN limits vary greatly, based on the shape of the theoretical SRN spectrum at energies that are below SK's SRN analysis threshold.  To remove this strong model dependence, a limit was set for $\rm E_\nu > 19.3\,MeV$.  In this region, all six models have similar energy spectrum shapes, and so an experimental limit that is insensitive to the choice of model can be obtained as follows:

\begin{equation}
\label{e:ind}
F_{ins} = F \times \frac{\int_{\rm 19.3\, MeV}^{\infty}{f(E_\nu)dE_\nu}}{\int_{0}^{\infty}{f(E_\nu)dE_\nu}}
\end{equation}

Flux limits in this energy region were the same for all models considered: $1.2\; \bar{\nu}_e$~cm$^{-2}$~s$^{-1}$.  Previously, the best limit on the SRN flux in this region was set using 357 days of data at Kamiokande-II~\cite{zhang}.  This limit was $226\; \bar{\nu}_e$~cm$^{-2}$~s$^{-1}$; the current SK limit is two orders of magnitude lower.


In summary, a search was conducted at \SK\ to detect the diffuse signal of $\bar{\nu}_e$ from all previous core-collapse supernovae.  No appreciable signal was detected in 1496 days of SK data.  Using various models, 90\% C.L. limits were set on the total SRN flux.  A limit of $1.2\; \bar{\nu}_e$~cm$^{-2}$~s$^{-1}$ was set for the SRN flux above a threshold of $E_{\nu} > 19.3$~MeV.  These results are more than an order of magnitude better than previous limits; some theories regarding the supernova rate in the universe can be constrained or rejected by these limits.


The authors gratefully acknowledge the cooperation of the Kamioka Mining and Smelting Company.  The \SK\ detector was built and operated with funds provided by the Japanese Ministry of Education, Culture, Sports, Science, and Technology; the U.S. Department of Energy; and the U.S. National Science Foundation.




\begin{thebibliography}{99}
\bibitem{hartmann}
D.~H.~Hartmann and S.~E.~Woosley, Astropart. Phys. {\bf 7}, 137 (1997).   The ``NC'' model is used here, which was selected by the authors as their standard model.
\end{thebibliography}
\end{document}